# WHAT DO AI/ML PRACTITIONERS THINK ABOUT AI/ML BIAS?

Aastha Pant, Monash University, Melbourne

Rashina Hoda, Monash University, Melbourne

Burak Turhan, University of Oulu, Finland

Chakkrit Tantithamthavorn, Monash University, Melbourne

*Abstract*— AI leaders and companies have much to offer to AI/ML practitioners to support them in addressing and mitigating biases in the AI/ML systems they develop. AI/ML practitioners need to receive the necessary resources and support from experts to develop unbiased AI/ML systems. However, our studies have revealed a discrepancy between practitioners' understanding of 'AI/ML bias' and the definitions of tech companies and researchers. This indicates a misalignment that needs addressing. Efforts should be made to match practitioners' understanding of AI/ML bias with the definitions developed by tech companies and researchers. These efforts could yield a significant return on investment by aiding AI/ML practitioners in developing unbiased AI/ML systems.

The widespread adoption of AI technology has brought about heightened ethical concerns, particularly regarding bias. Instances of bias have emerged in AI/ML applications, such as Amazon's recruitment tools favoring male candidates over females [1], Google's ML algorithms linking males more frequently with Science, Technology, Engineering, and Mathematics (STEM) careers [2], risk assessment algorithms displaying significant biases against African Americans, resulting in higher error rates in predicting future criminal behavior [3] and so on. These incidents impact software users diversely and may carry serious consequences.

---

Upon getting ethics approval (Reference Number: 38991), we conducted semi-structured interviews with 22 AI/ML practitioners in two rounds. The interviews were conducted using Zoom, audio-recorded and transcribed. The interview data was analysed using the Socio-Technical Grounded Theory (STGT) for data analysis [4]. Data collection and analysis were done iteratively. We also wrote memos to record our insights and reflections during the open coding process.

Most participants were from Australia (13 out of 22), followed by Nepal (3), and one each from the USA, India, Japan, Israel, Thailand, and Vietnam. 17 were men, and 5 were women. 9 had 2-5 years of AI/ML development experience, and 2 had less than 1 year. The majority held job titles as 'ML Engineers' (9) and 'AI engineers' (6), with most (19) involved in 'data cleaning' activities.

Table 1: Our Methodology

But how do AI/ML practitioners engage with these issues and how does that impact the way we deal with these issues? There is a recognised need for further research to investigate the perspectives and experiences of AI practitioners regarding AI ethics [5], specifically AI/ML bias [6].

Our empirical analysis (see Table 1) shows that AI/ML practitioners primarily understood and described 'AI/ML bias' in two ways: *(i) In terms of output of AI/ML,* and *(ii) In terms of factors leading to AI/ML bias*.

**A. In terms of output of AI/ML:** The participants described 'AI/ML bias' in terms of the nature of the output that AI/ML produces, with the system's preference as the underlying concept.

*1) System's preference:* A majority of the participants [P1, P3, P4, P5, P6, P7, P8, P11, P14, P18, P19, and P20] described the concept of 'AI/ML bias' as the system's preference or favoring of one thing over the other.

> *"I understand that ML bias refers to a bias in an ML model that favors one thing over others."* P1, AI Engineer, Nepal

**B. In terms of factors leading to AI/ML bias:** Several participants described 'AI/ML bias' in terms of two categories of factors contributing to the biases during AI/ML development including (i) Inappropriate dataset usage and (ii) Inappropriate algorithm selection.

*1) Inappropriate dataset usage:* A majority of the participants [P1, P2, P3, P4, P5, P6, P8, P9, P10, P12, P13, P14, P15, P16, P17, P18, P21, P22] described the term 'AI/ML bias' in terms of the use of inappropriate datasets during the development of such systems.

> *"If you want to produce a model that can generalise, then you need to have a good distribution of geographical sites. If you have a dataset that has been collected from only some geographic localities then, that won't work."* - P17, ML Engineer, Australia

*2) Inappropriate algorithm selection:* Participants [P11 and P22] discussed AI/ML bias concerning the use or selection of inappropriate algorithms in the development phase of these systems.

> *"If you train a model with imbalanced datasets, it makes the model biased. It can also be caused by the use of poor algorithms along with data."* - P22, Data Scientist, USA

AI/ML practitioners also discussed the *people* and *resources* that help them mitigate biases during AI/ML development.

**A. People:** Participants identified four distinct groups who could assist them in mitigating biases during the development of AI/ML : (i) Domain experts as essential, (ii) Team members as valuable allies, (iii) AI/ML practitioners themselves, and (iv) Users through their feedback help.

*1) Domain experts as essential:* Participants [P1, P3, P4, P7, P9, P10, P17] reported that domain experts play an important role in helping them mitigate biases in the AI/ML they develop. They emphasised that domain experts provide crucial domain knowledge, enhancing awareness of biases inherent in the AI/ML being developed.

> *"I don't have any medical background. Sometimes, some factors or features I never thought about could cause bias. So we have other advisors like domain experts, from other universities, doctors, and professors. So we ask them if we have any doubts."* - P7, AI Engineer, Australia

*2) Team members as valuable allies:* Many participants [P1, P2, P4, P5, P8, P18, P19, P20] identified key team members within their organisations like legal teams, senior engineers, and testing teams as valuable allies in mitigating biases in AI/ML development. They highlighted that these team members could provide guidance or intervene actively, assisting practitioners in recognising biases within the AI/ML they develop.

> *"After training our model, we do some set of tests on the model to get the validation and we explain to the AI ethics team, (we have an internal AI ethics team). So we run all these parameters along with explanations past them. They have to do the initial validation."* - P5, AI Engineer, Australia

*3) AI/ML practitioners themselves:* Participants [P4, P6, and P11] highlighted the role of practitioners in mitigating bias in AI/ML. They emphasised that adhering to best practices during the development process and possessing good knowledge and skills significantly aids in bias mitigation in AI/ML.

> *"In the ML pipeline building, there are certain best practices that everyone is supposed to follow. And that has to be religiously followed so that any sort of bias can be addressed there."* - P4, Data Scientist, India

*4) Users through their feedback help:* [P9 and P12] emphasised the significant role of users in helping them mitigate biases in the AI/ML they develop. They stated that user feedback could be valuable in recognising different types of biases within their systems.

> *"We could always get the end users' feedback as well on the model we develop."* - P9, Data Scientist, Australia

**B. Resources:** AI/ML practitioners discussed two types of resources that aid them in mitigating biases during AI/ML development including: (i) Technological resources and (ii) Educational resources.

*1) Technological resources:* Several participants [P1, P2, P6, P7, P8, and P11] reported how technological resources such as GitHub, ChatGPT, blogs, and Stack Overflow, help them in mitigating biases during AI/ML development.

> *"I just type the problem in Google and go to some comments. People have some GitHub repos or so and I get information on biases and how they solved the issues."* - P6, Data Scientist, Australia

*2) Educational resources:* Participants [P5, P18, and P21] highlighted the role played by educational resources in helping them mitigate biases during AI/ML development. They emphasised the importance of research papers, textbooks, and workshops to educate them on different aspects of addressing bias-related issues during AI/ML development.

> *"I think we can have some workshops so that we can improve our awareness about this AI biases stuff."* - P18, ML Engineer, Australia

Figure 1 shows the key findings of our study.

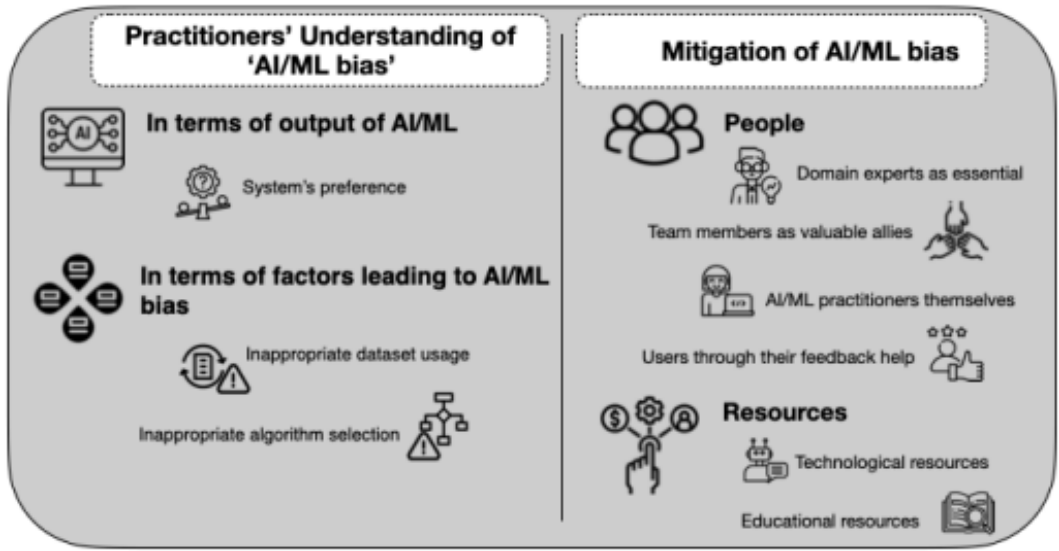

Fig 1. Key Findings of our study.

## RECOMMENDATIONS FOR AI LEADERS AND COMPANIES

Based on our findings and insights derived from *memoing* [4], we recommend the following to leaders and companies employing or engaging with AI/ML practitioners:

- *Foster interdisciplinary collaboration:* A majority of AI/ML practitioners in our study indicated that different team members could assist them with bias-related issues during AI/ML development. AI companies should make explicit attempts to foster collaboration between AI/ML practitioners and other teams like legal teams, senior engineering, and testing teams [7]. This collaboration can be facilitated through clear communication channels between teams and regular cross-team meetups and communities of practice.
- *Encourage collaboration with domain experts:* We found that domain experts are vital in assisting AI/ML practitioners in mitigating biases in AI/ML development by providing them with their specialised knowledge of the domain [7]. AI/ML practitioners should be provided the opportunities and encouraged to collaborate with domain experts throughout the AI/ML development life-cycle. This collaboration could begin from the problem definition phase to the deployment and monitoring phase to ensure that domain-specific knowledge is integrated at every step, reducing the risk of biased outcomes.
- *Facilitate workshops and training sessions:* Our findings show that AI/ML practitioners find educational resources like workshops, research papers, and textbooks useful to address their queries on AI/ML bias. These resources provide information and insights for navigating bias-related challenges during development. We recommend organising workshops and training sessions within the company to educate practitioners on bias mitigation in AI/ML development. Experts from academia and industry can be invited to share their insights, methodologies, and practical experiences in addressing bias-related challenges. Our previous empirical study on 'AI ethics' also highlighted the need for training programs and formal education for AI practitioners to incorporate ethical considerations during AI development [8].

- *Develop a definition of 'AI/ML bias' collaborating with AI/ML practitioners*: Big tech companies like IBM [9] have defined 'AI bias' as *"the occurrence of biased results due to human biases that skew the original training data or AI algorithm—leading to distorted outputs and potentially harmful outcomes."* Our results show that AI/ML practitioners may describe the term 'AI/ML bias' slightly differently from how companies define it. In our study, AI practitioners described 'AI/ML bias' not only *in terms of the output of AI/ML systems*, aligning with definitions from companies like IBM, but also *in terms of the contributing factors leading to AI/ML bias*. These contributing factors are often not mentioned in the definitions of tech companies like IBM. This blindspot could result in the development of biased AI/ML systems, as the definitions and guidelines meant to guide practitioners in developing unbiased AI/ML may be misaligned. Therefore, we recommend that AI companies and policy-making bodies consider developing guidelines and definitions for 'AI/ML bias' by incorporating AI/ML practitioners' more comprehensive insights and understanding of AI/ML bias.

**ACKNOWLEDGEMENTS:** This work involved human subjects in its research. Ethical approval (Reference Number: 38991) was granted by Monash University to conduct interviews with AI/ML practitioners. Consent from all the participants in our research was obtained, and all data were anonymised. We would like to thank all the participants of our study.